\documentclass[twocolumn,showpacs,preprintnumbers,pre]{revtex4}
\usepackage{color}
\usepackage{epsfig}
\usepackage{amsmath}
\usepackage{amssymb}
\usepackage{verbatim}
\usepackage{graphicx}
\usepackage{natbib}
\begin{document}

\title{Normal Modes in Model Jammed Systems in Three Dimensions}

\author{Leonardo E. Silbert$^{1}$}

\author{Andrea J. Liu$^{2}$}

\author{Sidney R. Nagel$^{3}$}

\affiliation{$^{1}$Department of Physics, Southern Illinois
  University, Carbondale, IL 62901, U.S.A.}

\affiliation{$^{2}$Department of Physics and Astronomy, University of
  Pennsylvania, Philadelphia, PA 19104, U.S.A.}

\affiliation{$^{3}$James Franck Institute, University of Chicago,
  Chicago, IL 60637, U.S.A.}

\date{\today}

\begin{abstract}

  Vibrational spectra and normal modes of mechanically stable particle
  packings in three dimensions are analyzed over a range of
  compressions, from near the jamming transition, where the packings
  lose their rigidity, to far above it. At high frequency, the normal
  modes are localized at all compressions. At low frequency, the
  nature of the modes depends somewhat on compression.  At large
  compressions, far from the transition, the lowest-frequency normal
  modes have some plane-wave character, though less than one would
  expect for a crystalline or isotropic solid. At low compressions
  near the jamming transition, the lowest-frequency modes are neither
  plane-wave-like nor localized.  We characterize these differences,
  highlighting the unusual dispersion behavior that emerges for
  marginally jammed solids.

\end{abstract}


\maketitle

It is well recognized that the high-frequency vibrations in amorphous
materials are strikingly different from those in crystals. In glasses
and other amorphous solids, the highest-frequency normal modes are
localized in space, while in crystals they are extended excitations
\cite{bell1,sompolinsky1,nagel11}. It is also appreciated that even at
low frequencies the normal modes of disordered systems can be
dramatically different from the long-wavelength plane waves found in
ordered materials. The vibrational spectra of amorphous solids are
characterized by ``boson peaks'' -- extra low-frequency modes beyond
the long-wavelength plane-wave phonons found in crystals. The
anomalous modes that fall within the boson peak are believed to be
responsible for the unusual low-temperature properties of glasses,
such as the plateau in the thermal conductivity \cite{phillips1}.

Nowhere are the excess low-frequency excitations more apparent than in
a marginally jammed solid, in which a system of particles is
compressed to the point where they first begin to touch and form a
rigid structure
\cite{durian5,ohern2,ohern3,leo14,wyart1,wyart2,durian6,hecke7,behringer9}.
In this system, the density of normal modes, instead of vanishing as
the frequency is lowered towards zero, as in a crystal, remains
constant as shown in Fig. 1(a). This leads to an excess in the density
of states that {\it diverges} at zero frequency.  This divergence has
been interpreted as a boson peak \cite{leo14}.  Thus, one might expect
the marginally jammed solid to provide the clearest window into the
anomalous low-frequency normal modes characteristic of all amorphous
solids.

Previously, we have found that the characteristic frequency and size
of the boson peak can be tuned systematically by compressing the
marginally jammed solid to higher packing fractions \cite{leo14}. In
the present paper we analyze the structure of the normal modes of
disordered packings in three dimensions ($3D$) in the marginally
jammed state and at compressions above this state. The system we study
here is identical to the one used previously to calculate the density
of vibrational states \cite{ohern3,leo14,leo12}. We simulate a $3D$
system of $N$ ($1024 \le N \le 10000$) monodisperse soft-spheres of
mass $m$ and diameter $\sigma$ in cubic simulation cells employing
periodic boundary conditions. The particles interact via a
finite-range, purely repulsive, harmonic potential:
\begin{equation}
V(r)= 
\left\{
  \begin{array}{lrr} 
    V_{0}(1-r/\sigma)^2 && r < \sigma
    \\0 && r \geq \sigma 
  \end{array}
\right\}
\label{eq1}
\end{equation}
where $r$ is the center-to-center distance between two particles.
Length and time are measured in units of $\sigma$ and
$(md^{2}/V_{0})^{1/2}$ respectively.  We initially place N particles
at random in a cubic box of linear dimension $L$.  This corresponds to
a $T=\infty$ configuration.  We use conjugate-gradient energy
minimization \cite{recipes} in order to obtain $T = 0$ configurations.
In order to show visualizations of the normal modes, we have also
studied $N=10000$, bidisperse, 50:50 mixtures of harmonic discs with a
diameter ratio of $1.4$ in $2D$.

The onset of jamming in the limit of large $N$ occurs at a packing
fraction, $\phi_{c}^\infty = 0.639 \pm 0.003$, and is characterized by
the onset of a nonzero pressure. We determine $\phi_{c}$ for each of
our finite-system initial configurations by incrementally compressing
(decompressing) until the pressure just becomes nonzero (just reaches
zero). We then compress the system to obtain zero-temperature
compressed configurations at controlled values of $\phi-\phi_c$. For
each of these configurations, we compute and diagonalize the dynamical
matrix \cite{ashcroft1}. The eigenvalues and eigenmodes of this matrix
are respectively the squared frequencies, $\omega^{2}$, of the normal
modes of vibration and the corresponding polarizations ${\bf
  e}_{i_{\omega}}$ of each particle $i$ in the normal mode of
frequency $\omega$.

\begin{figure}[h]
  \includegraphics[width=8cm,height=4cm]{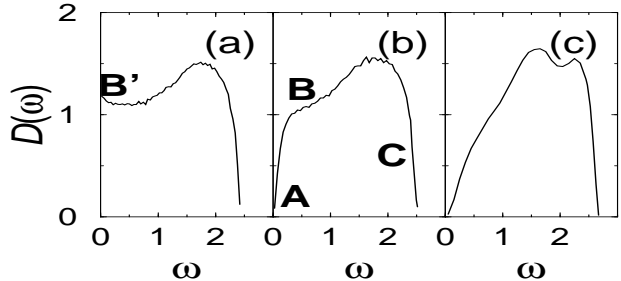}
  \caption{Density of states $\mathcal{D}(\omega)$ of $3D$ jammed
    packings with $N=1024$ monodisperse spheres at three different
    compressions: (a) $\Delta\phi = 10^{-6}$, (b) $\Delta\phi =
    10^{-2}$, and (c) $\Delta\phi = 10^{-1}$. In (a) we identify
    regime B', and in (b) we identify the regimes A,B, and C,
    discussed in the text.}
  \label{fig1}
\end{figure}

In a previous paper \cite{leo14}, we analyzed the density of
vibrational states $D(\omega)$, of configurations above the jamming
threshold, for systems at $\phi>\phi_c$, and found three
characteristic regimes, as labeled in Fig.~\ref{fig1}(b). In regime A,
$D(\omega)$ decreases towards zero as $\omega \rightarrow 0$. In
regime B (including B' in Fig.~\ref{fig1}(a)), $D(\omega)$ is
approximately constant, very different than for crystals.  Finally, in
regime C at high frequencies, $D(\omega)$ decreases with increasing
frequency. Figure \ref{fig1}(a-c) show how the different regimes shift
with increasing compression. At high $\Delta \phi \equiv
\phi-\phi_{c}$, regime A extends to fairly high frequencies and regime
B is small. As the system is decompressed towards the marginally
jammed solid at $\Delta \phi=0$, regime A shrinks and regime B grows,
extending all the way down to zero frequency, as indicated by B' in
Fig.~\ref{fig1}(a), while regime C remains approximately the same. The
growth of regime B at the expense of regime A with decreasing
compression signals the proliferation of anomalous low-frequency
modes. These are the modes whose structure we wish to understand.

\begin{figure}[h]
  \includegraphics[width=3.4cm]{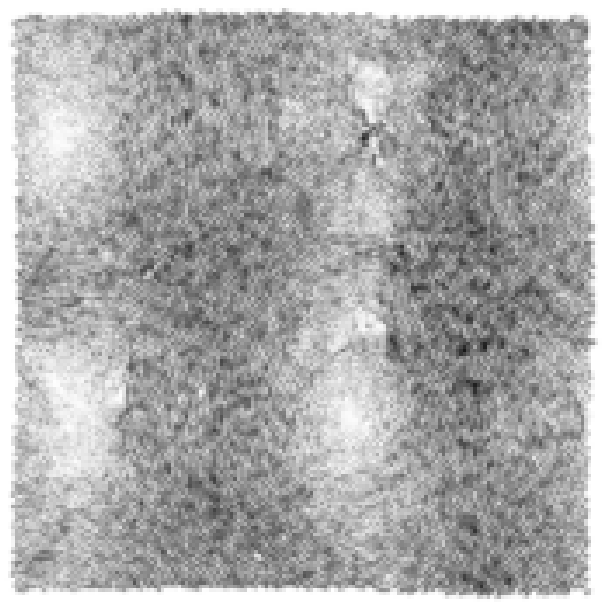}
  \includegraphics[width=3.4cm]{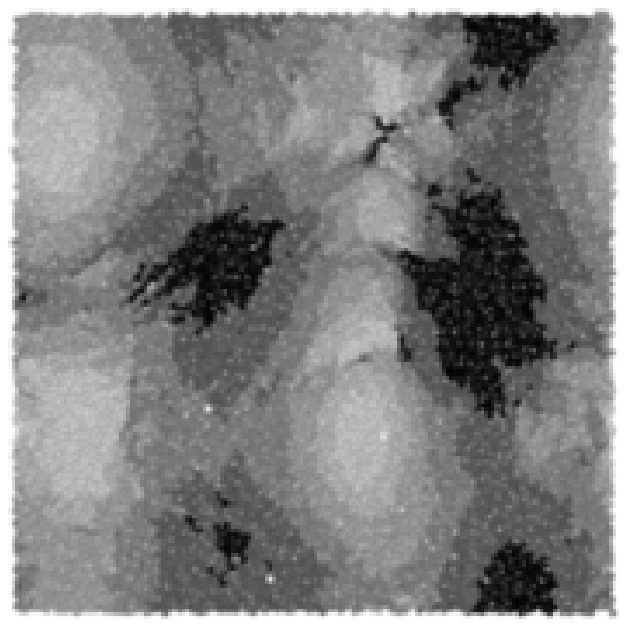}\\
  \includegraphics[width=3.4cm]{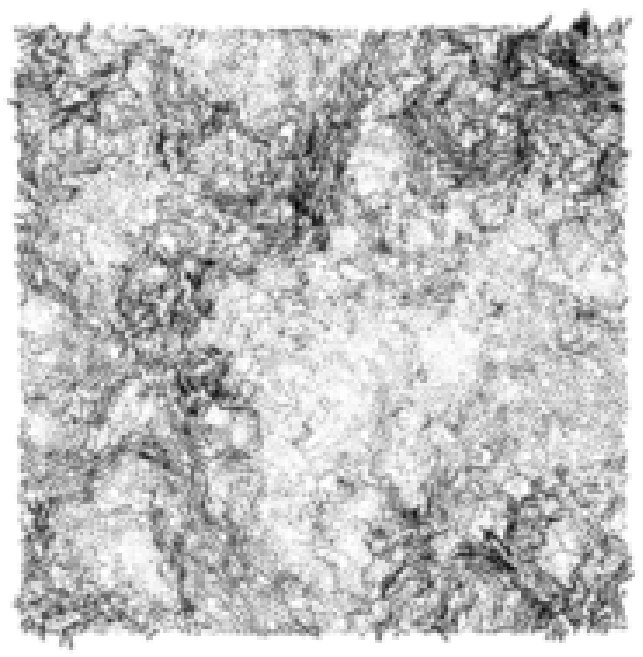}
  \includegraphics[width=3.4cm]{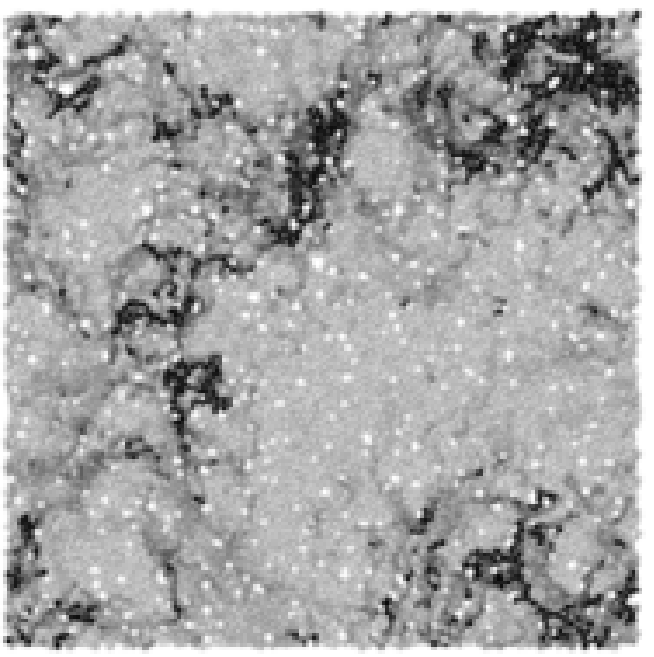}\\
  \includegraphics[width=3.4cm]{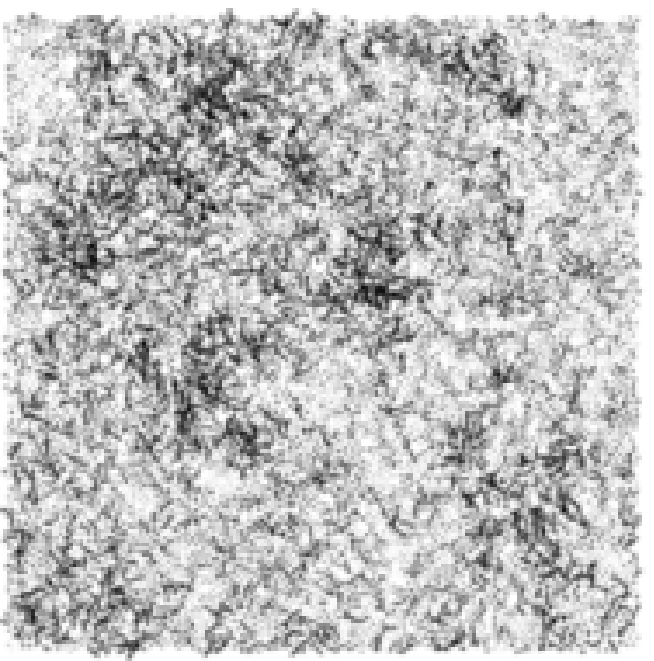}
  \includegraphics[width=3.4cm]{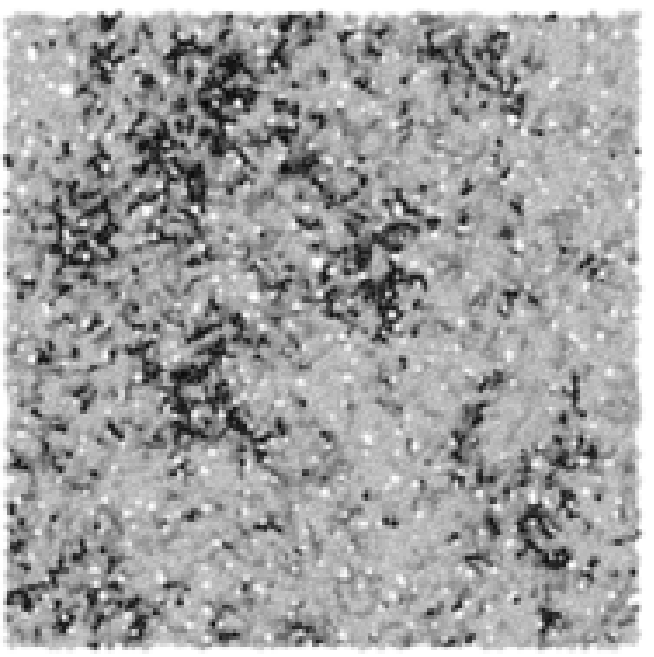}\\
  \includegraphics[width=3.4cm]{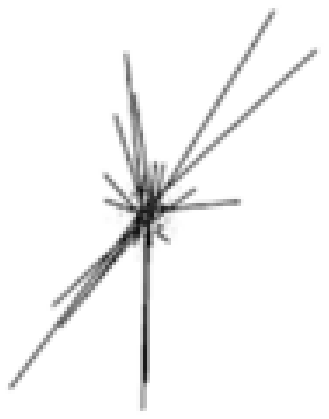}
  \includegraphics[width=3.4cm]{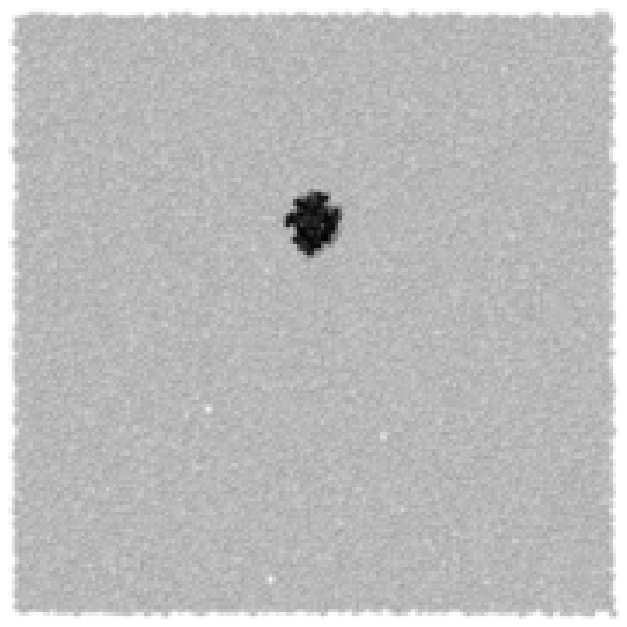}\\
  \hfill\includegraphics[width=4cm]{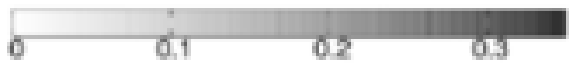}
  \caption{Normal modes of a $2D$ jammed packing of a 50:50 mix of
    $N=10000$ bidisperse disks, with size ratio $1:1.4$, interacting
    with the potential defined in Eq.~\ref{eq1}. These figures
    correspond to the regimes identified in Fig.~\ref{fig1}(a),(b)
    characterized by $(\Delta\phi,\omega)$. Top row panels: Regime A
    $(1 \times 10^{-1}, 1.76 \times 10^{-2})$. Second row: Regime B'
    $(1 \times 10^{-6}, 3.35 \times 10^{-4})$. Third row: Regime B $(1
    \times 10^{-6}, 0.3)$. Bottom row: Regime C $(1 \times 10^{-1},
    2.30)$. Left panels: black lines represent the amplitude and
    direction of the particle vibrations in that mode. Right panels:
    particles are shaded according to the magnitude of their
    polarization vector. The scale bar indicates that darker shading
    corresponds to a larger ratio of the amplitude to the maximum
    amplitude of particle displacement in that mode. Particles with no
    contact neighbors are not shown.}
  \label{fig2}
\end{figure}

In order to visualize the nature of the modes, we turn to the
bidisperse 2D system.  Fig.~\ref{fig2} shows typical normal modes from
regimes A, B', B, and C, defined in Fig.~\ref{fig1}(a) and (b). In the
left panels, the polarization vector for each particle is shown, while
in the right panel, each particle is shaded according to the magnitude
of its polarization vector. The mode from regime $A$,
Fig.~\ref{fig2}(top), appears to have some plane-wave-like character,
although contributions from several different wavevectors are readily
apparent. The high-frequency mode corresponding to regime C,
Fig.~\ref{fig2}(bottom) is quite localized. This particular mode is
representative of the high-frequency modes at all values of
$\Delta\phi$; visualizations for different $\Delta\phi$ are
indistinguishable from that shown in Fig.~\ref{fig2}(bottom).

The modes from regimes B' and B, Fig.~\ref{fig2}(middle), are neither
plane-wave-like nor localized. The right panels more clearly reveal
the filamentary nature of the extended vibrational modes in regimes B'
and B. Here we point out that these visualizations already suggest
some subtle differences between regimes B' and B which we quantify
below (see Fig.~\ref{fig7} and related discussion).

\begin{figure}[h] 
  \includegraphics[width=8cm]{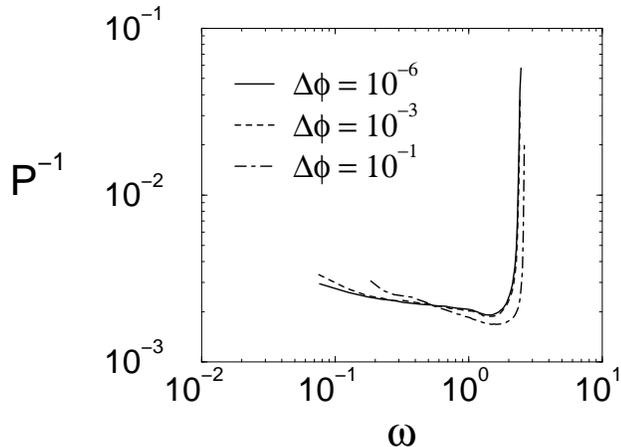}
  \caption{Inverse participation ratio, $P^{-1}$, for $3D$ jammed
    packings of $N=1024$ monodisperse spheres, at three compressions.}
  \label{fig3}
\end{figure}

Fig.~\ref{fig2} suggests that the modes from both regimes $A$ and $B$
are extended. To analyze directly how extended each mode is, we
calculate the inverse participation ratio,
\begin{equation}
  P^{-1} = \frac{\sum_{i=1}^{N} \left|{\bf
        e}_{i_{\omega}}{\bf \cdot e}_{i_{\omega}}\right|^{2}}{\left|\sum_{i=1}^{N} {\bf
        e}_{i_{\omega}}{\bf \cdot e}_{i_{\omega}}\right|^{2}}.
  \label{eq2}
\end{equation}
Here ${\bf e}_{i_{\omega}}$ is the polarization vector of particle $i$
in the mode $\omega$. We show the results for $\Delta \phi = 10^{-6}$,
$\Delta \phi = 10^{-3}$, and $\Delta \phi = 10^{-1}$ in
Fig.~\ref{fig3}.  We find that on a log-log scale, the participation
ratio looks quite similar for values of $\Delta \phi$ up to at least
$\Delta \phi \approx 10^{-3}$.  For high compressions ($\Delta \phi
=0.1$ and higher), $P^{-1}$ at the very highest frequencies (regime C)
appears to decrease with increasing compression, but $P^{-1}$ is still
large in regime C at all compressions. Thus, the modes in regime C are
localized at all compressions.  At lower frequencies, however, up to
$\omega \approx 2$, we find $P^{-1}<<1$, indicating that the modes are
extended over the size of the system.  Thus, modes in regimes A and B
are extended, while those in regime C are localized.

\begin{figure}[h] 
  \includegraphics[width=8cm]{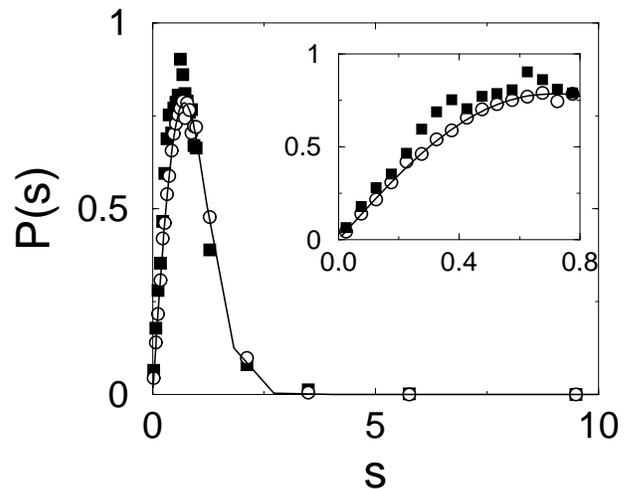}
  \caption{Distribution, $P(s)$, of level spacings, $s$, normalized by
    the the average level spacing, at $\Delta\phi = 10^{-1}$
    ($\blacksquare$) and $10^{-6}$ ($\circ$) in 3D. Inset shows the linear
    behavior at small $s$. The line is a fit to the Wigner-Dyson
    function of Eq.~\ref{eq3}.}
  \label{fig4}
\end{figure}

For these jammed, mechanically stable packings, there exist local
correlations in the force constants that constitute the dynamical
matrix: The forces around each and every particle must be locally
balanced.  Upon analyzing the spacings $\Delta\omega=\omega_{j+1} -
\omega_{j}$, between successive normal mode frequencies, we find
level-repulsion in the distribution of level spacings, $P(s)$, where
we define $s = \frac{\Delta\omega}{<\Delta\omega>}$, as the level
spacing normalized by the average $<\Delta\omega>$. The distributions
shown in Fig.~\ref{fig4} show little dependence on distance to the
jamming threshold, and are described quite well by the Wigner-Dyson
distribution \cite{mehta2},
\begin{equation}
  P(s) = \frac{\pi s}{2}e^{-\pi s^{2}/4}.
  \label{eq3}
\end{equation}
The distributions shown in Fig.~\ref{fig4} are peaked around the
average spacing and indicate level repulsion by the linear behavior at
small spacing. Thus, even though the dynamical matrix is sparse, due
to the short-range nature of the interaction potential, the level
spacings are not completely random, which would lead to a Poisson
distribution, nor are they completely correlated \cite{wim1}.

\begin{figure}[h]
  \includegraphics[width=8cm]{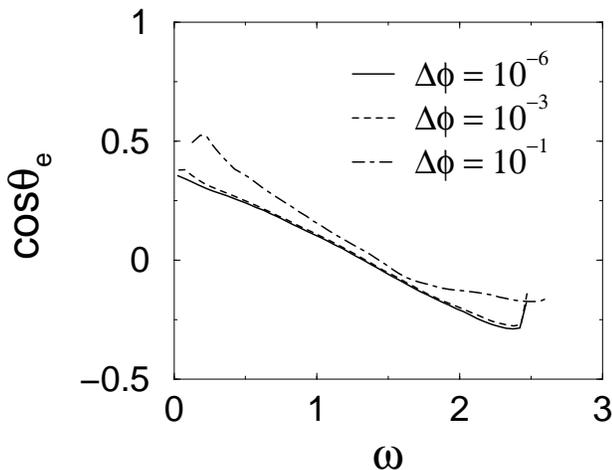}
  \caption {Correlation between particle displacements in a mode,
    measured using $\cos\theta_{e}$ defined by Eq.~\ref{eq4}, at three
    compressions for $3D$ jammed packings with $N=1024$ monodisperse
    spheres.}
  \label{fig5}
\end{figure}

Another way to characterize the modes is to look at local correlations
of the polarizations of neighboring particles. We calculate a
quantity, $\cos \theta_{e}$, that is similar to the phase quotient
parameter often probed in glasses \cite{allen2},
\begin{equation}
  \cos \theta_{e}(\omega) = \frac{1}{N_{\rm{pairs}}} \sum_{i,j}
  \hat{{\bf e}}_{i_{\omega}} \cdot \hat{{\bf e}}_{j_{\omega}}
  \label{eq4}
\end{equation}
where the sum only runs over the number of pairs of particles that
interact with each other, $N_{\rm pairs}$, and the normalized
polarization vector of particle $i$ associated with the normal mode of
frequency $\omega$, $\hat{\bf e}_{i_{\omega}}=\frac{{\bf
    e}_{i_{\omega}}}{|{\bf e}_{i_{\omega}}|}$. For a mode in which
every particle is vibrating in approximately the same direction, i.e.,
strongly correlated motion, one would expect $\cos \theta_e \approx
1$. Figure ~\ref{fig5} shows $\cos \theta_e$ as a function of
frequency at three values of $\Delta \phi$. Up to compressions of
order $\Delta \phi=10^{-3}$, the behavior of $\cos \theta_e$ with
$\omega$ is insensitive to compression.  Over that range of
compressions, $\cos \theta_e$ decreases linearly with frequency,
showing that in the low-frequency modes, particle displacements are
more correlated with their neighbors and in high frequency modes,
particle displacements are more anti-correlated with their neighbors.
Note that the frequency ranges of regimes A and B change appreciably
with $\Delta \phi$, while $\cos \theta_e(\omega)$ does not; this
suggests that the correlations are not noticeably different in the two
regimes.  At high compressions, the curve begins to show a kink at
approximately $\omega = 1.75$ (the boundary between regimes B and C).

\begin{figure}[h]
  \includegraphics[width=4cm]{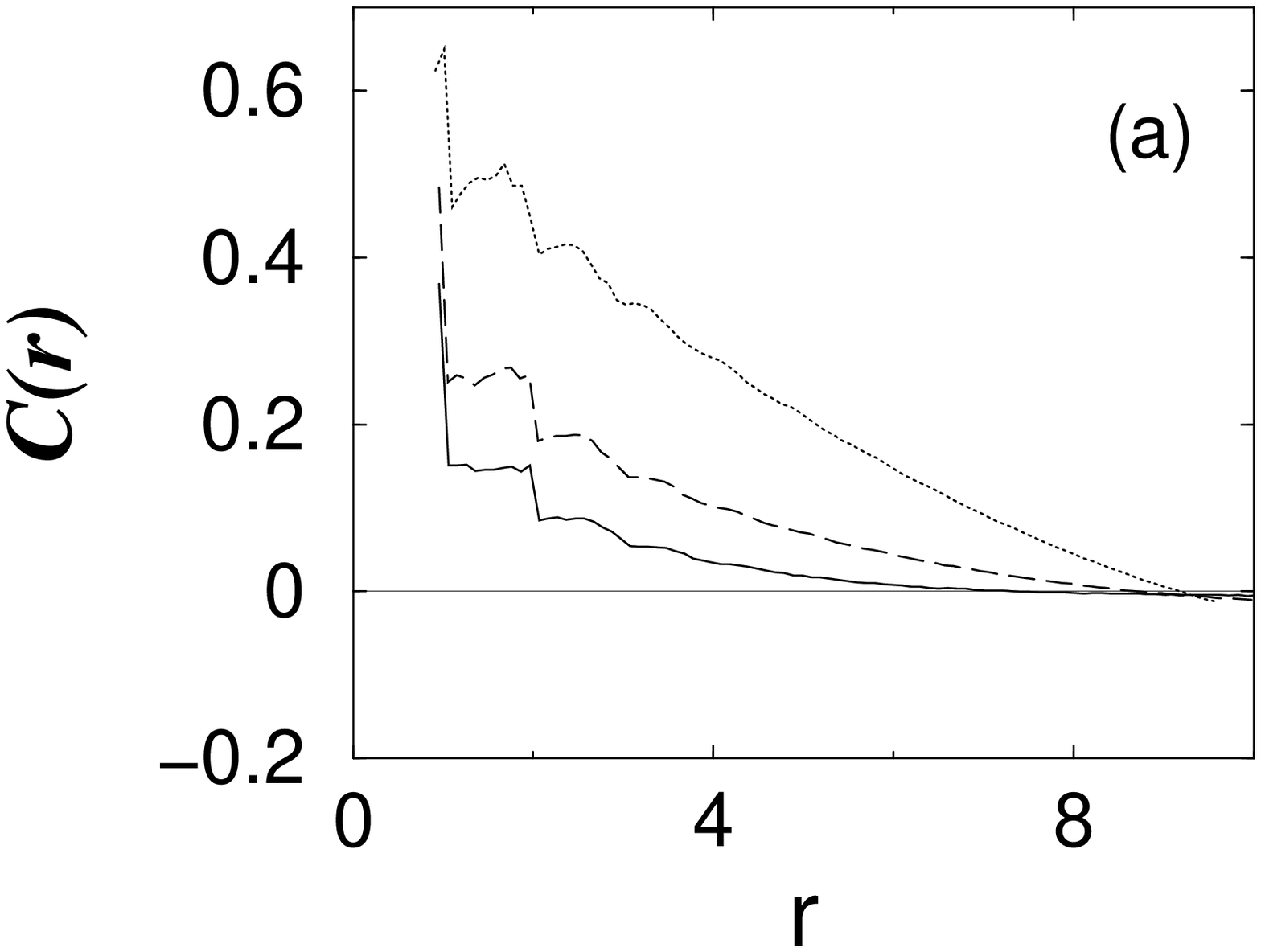}
  \includegraphics[width=4cm]{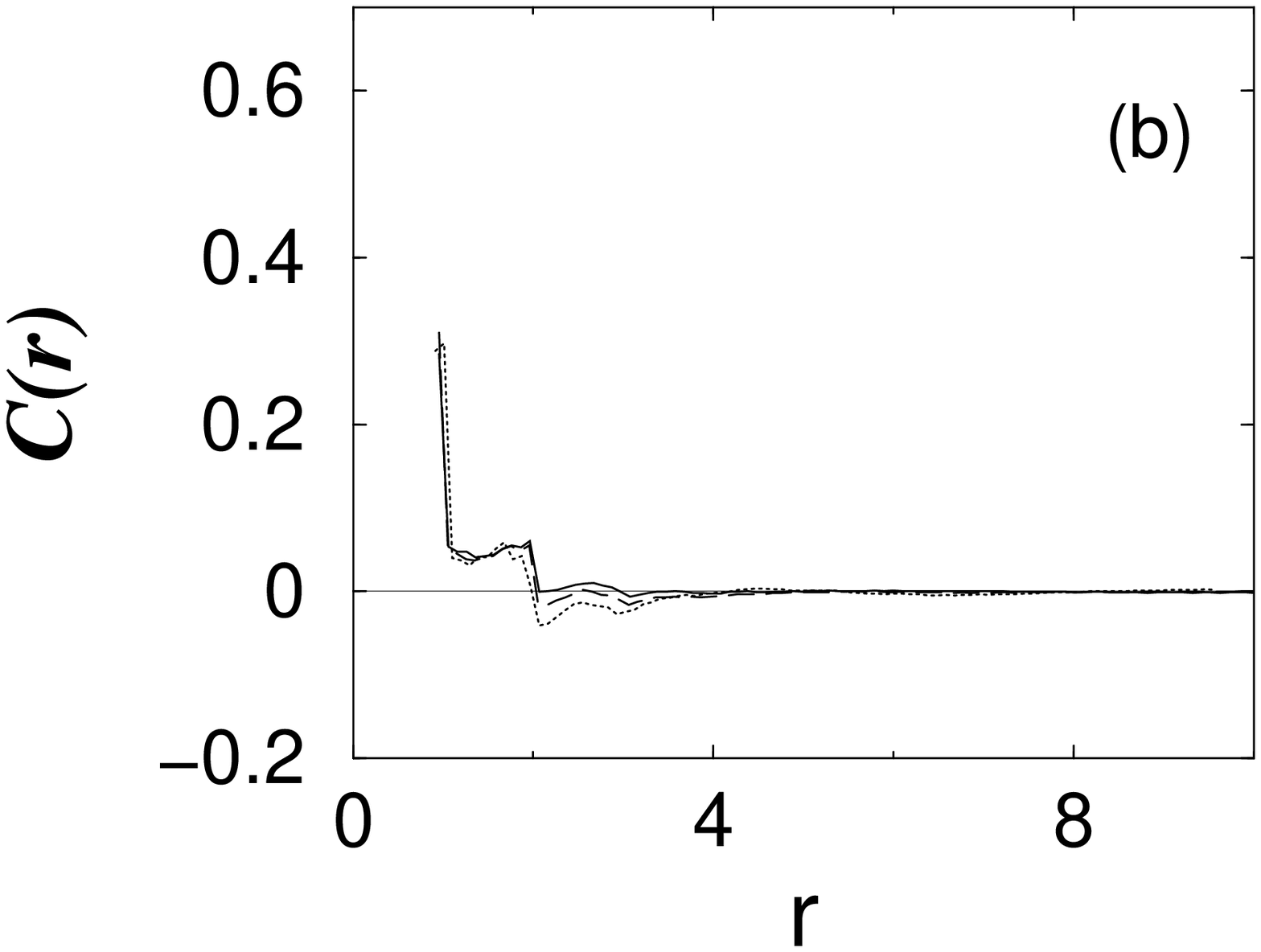}
  \caption{Spatial correlation function ${\mathcal C}(r)$ of particle
    vibrations at three values of $\Delta\phi = 10^{-6}$ (solid line),
    $10^{-2}$ (dashed), and $10^{-1}$ (dotted), for $N=10000$
    monodisperse spheres in $3D$. (a) Lowest-lying frequencies, and (b)
    intermediate frequencies.}
  \label{fig6}
\end{figure}

\begin{figure}[h]
  \includegraphics[width=8cm]{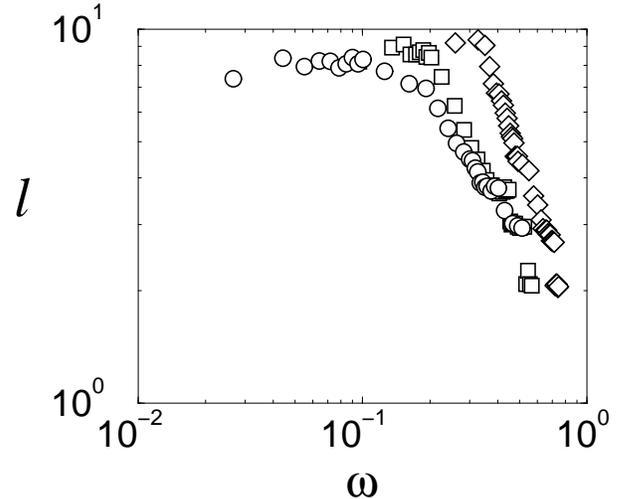}
  \caption{Dependence of the characteristic length scale $\ell$,
    defined in text, on the normal mode frequency $\omega$, at three
    values of $\Delta\phi = 10^{-6}$ ($\bigcirc$), $10^{-3}$
    ($\square$), and $10^{-1}$ ($\lozenge$), for $N=10000$
    monodisperse spheres in $3D$.}
  \label{fig7}
\end{figure}

As noted earlier (and by visual inspection of the $2D$ modes of
Fig.~\ref{fig2}), the low-frequency modes both near and far from the
jamming threshold appear to have somewhat different character. We have
argued that compressed systems away from the jamming threshold contain
low-frequency modes that are more plane-wave-like, whereas near the
jamming transition, these extended modes are very different from plane
waves. At intermediate and high frequencies the modes appear
relatively insensitive to packing fraction. In an effort to further
differentiate between the nature of the low-frequency modes we measure
the spatial extent of correlated vibrational motions. For each mode of
frequency $\omega$, we compute the correlation of polarization vectors
between particles $i$ and $j$,
\begin{equation}
  {\mathcal C}(r_{ij}) = \left< \hat{{\bf e}}({\bf r}_{i})\cdot\hat{{\bf e}}({\bf r}_{j})\right>.
  \label{eq5}
\end{equation}
In Fig.~\ref{fig6} we show ${\mathcal C}(r)$ for the lowest lying
frequency modes at three different compressions, $\Delta\phi =
10^{-6}$, $\Delta\phi=10^{-2}$, and $\Delta\phi=10^{-1}$, for our $3D$
jammed packings.

For the low-frequency modes we find somewhat stronger correlations at
higher compressions.  For the middle-to-high frequency range of the
vibrational spectrum, beyond regime A, the normal modes become
increasingly similar at different compressions. In regime C, the modes
are indistinguishable at different compressions.

All of the correlation functions in Fig.~\ref{fig6} decay
non-monotonically to zero and cross zero at some finite $r$. We define
the value of $r$ at which ${\mathcal C}(r)$ first crosses zero to be
$\ell (\omega)$.  In a pure plane wave, particle vibrations are
correlated and ${\mathcal C}(r)$ will cross zero at the scale of half
the wavelength, so for ordinary sound modes at low frequency, one
would expect $\ell(\omega) \propto 1/\omega$. In Fig.~\ref{fig7}, we
plot $\ell$ as a function of $\omega$ for different $\Delta\phi$ at
low-to-intermediate frequencies. For the system closest to the jamming
threshold, $\ell$ is approximately independent of $\omega$ at very low
frequencies.  Beyond this constant region, $\ell$ decreases with
increasing frequency, corresponding to moving along the plateau in the
density of states from regime B' to B \cite{footnote22}. As the system
moves further from the jamming threshold, i.e., as $\Delta\phi$
increases, the region of constant $\ell$ shrinks. Over the range of
frequencies where $\ell$ decreases, the characteristic length is
greater the further the system is from the jamming threshold. This is
to be expected as the modes contain more plane-wave character at
larger compressions. At slightly higher frequencies still, the curves
begin to overlap, so that the modes do indeed become practically
indistinguishable. The frequency at which this occurs coincides with
the point in the density of states where the plateau regions start to
merge (see Fig.~\ref{fig1} of Ref.~\onlinecite{leo14}). These data
suggest that the distinction between modes from regimes B' and B is
related to how extended the modes are. This is already evident from
the visualizations presented in Fig.~\ref{fig2}.

Another way to quantify the difference between the extended modes in
regimes A and B is through the Fourier transforms of the eigenmodes at
different frequencies $\omega$ throughout the spectrum.  Specifically,
we take the Fourier transform of the appropriate component, either
longitudinal or transverse, of the particle polarization vector ${\bf
  e}_{j}(\omega)$, of each particle $j$ \cite{nagel11,leo14}:
\begin{equation}
  \begin{array}{lrr} 
    f_{L}(k,\omega) &=& \left\langle \frac{1}{N}\left| \sum_{j} {\bf\hat{k}} \cdot
        {\bf e}_{j_{\omega}} \exp(\imath {\bf k} \cdot {\bf
          r}_{j})\right|^{2}\right\rangle,\\
    &&\\
    f_{T}(k,\omega)& =& \left\langle \frac{1}{N}\left|\sum_{j} {\bf\hat{k}} \wedge
        {\bf e}_{j_{\omega}} \exp(\imath {\bf k} \cdot {\bf
          r}_j)\right|^2\right\rangle.
  \end{array}
  \label{eq6}
\end{equation}

\begin{figure}[h] 
  \includegraphics[width=4cm]{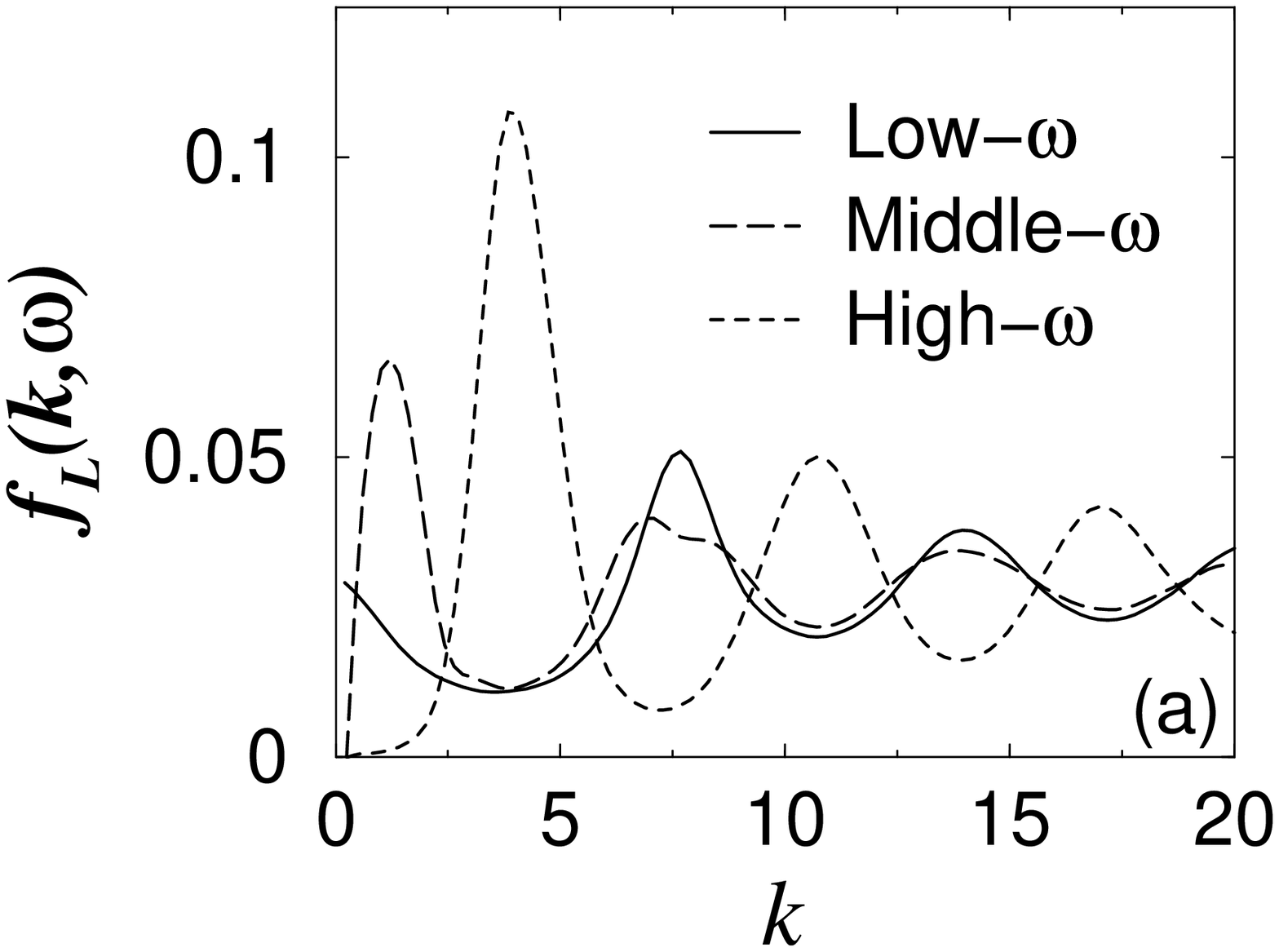}
  \includegraphics[width=4cm]{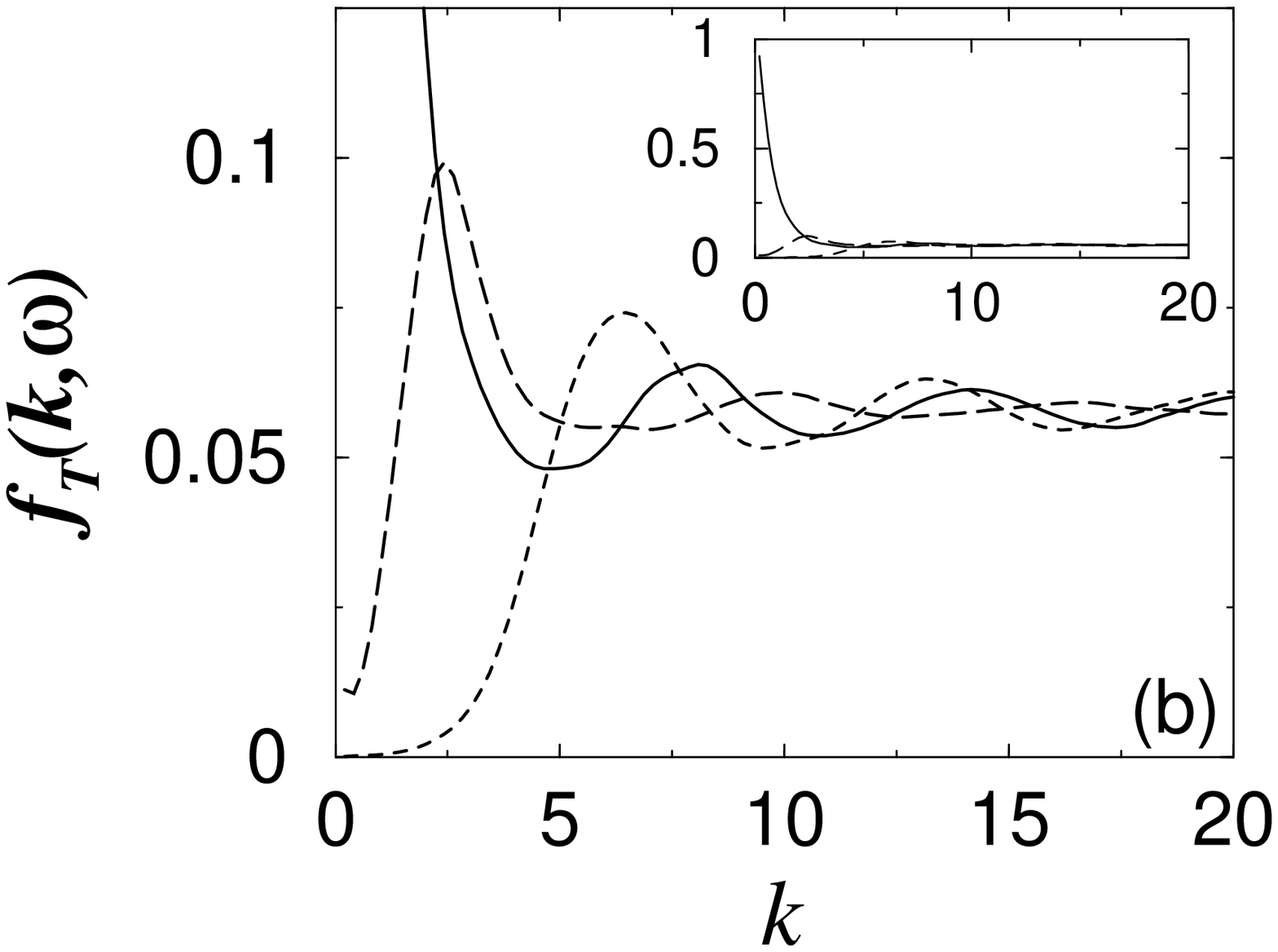}\\
  \includegraphics[width=4cm]{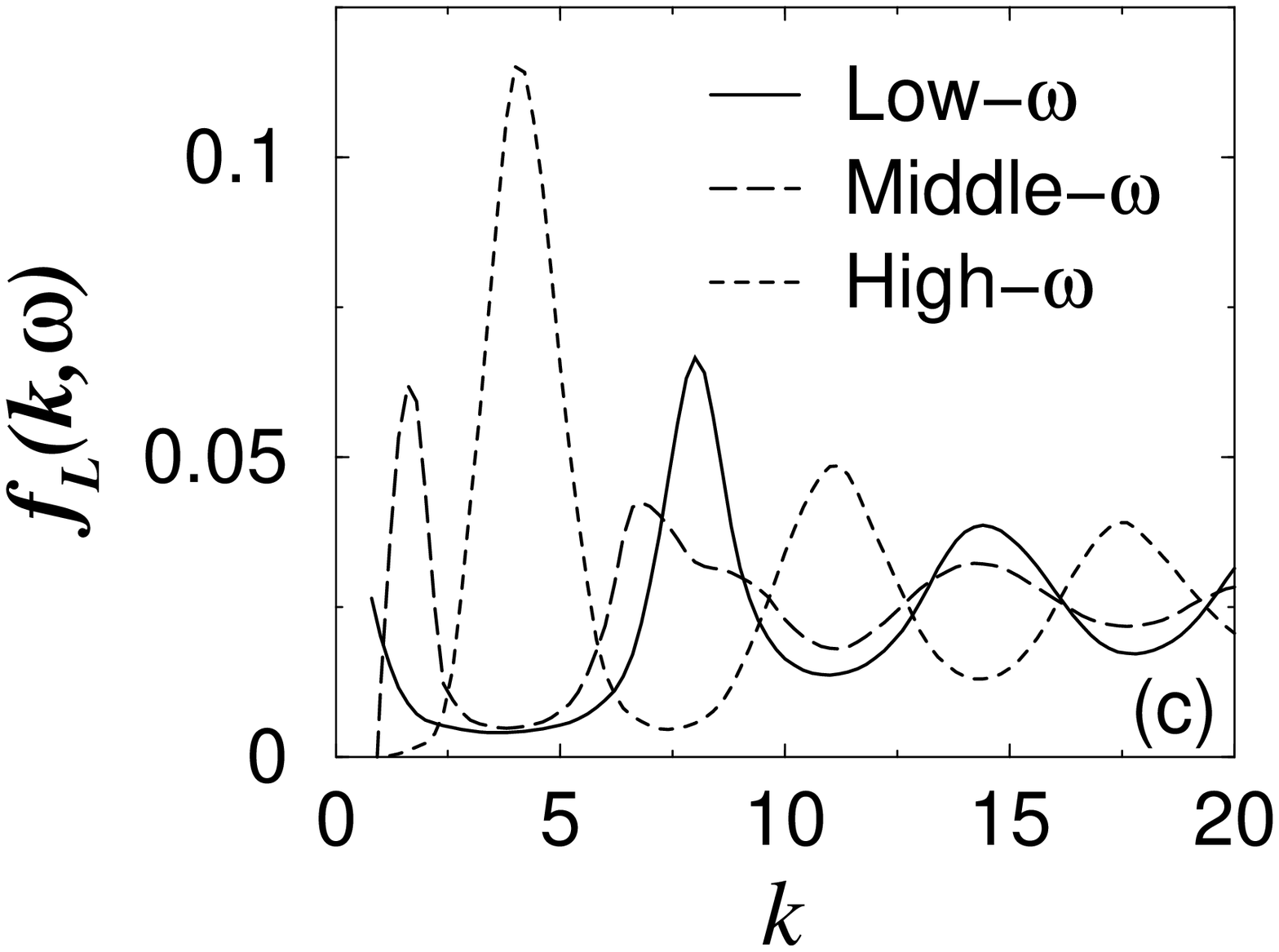}
  \includegraphics[width=4cm]{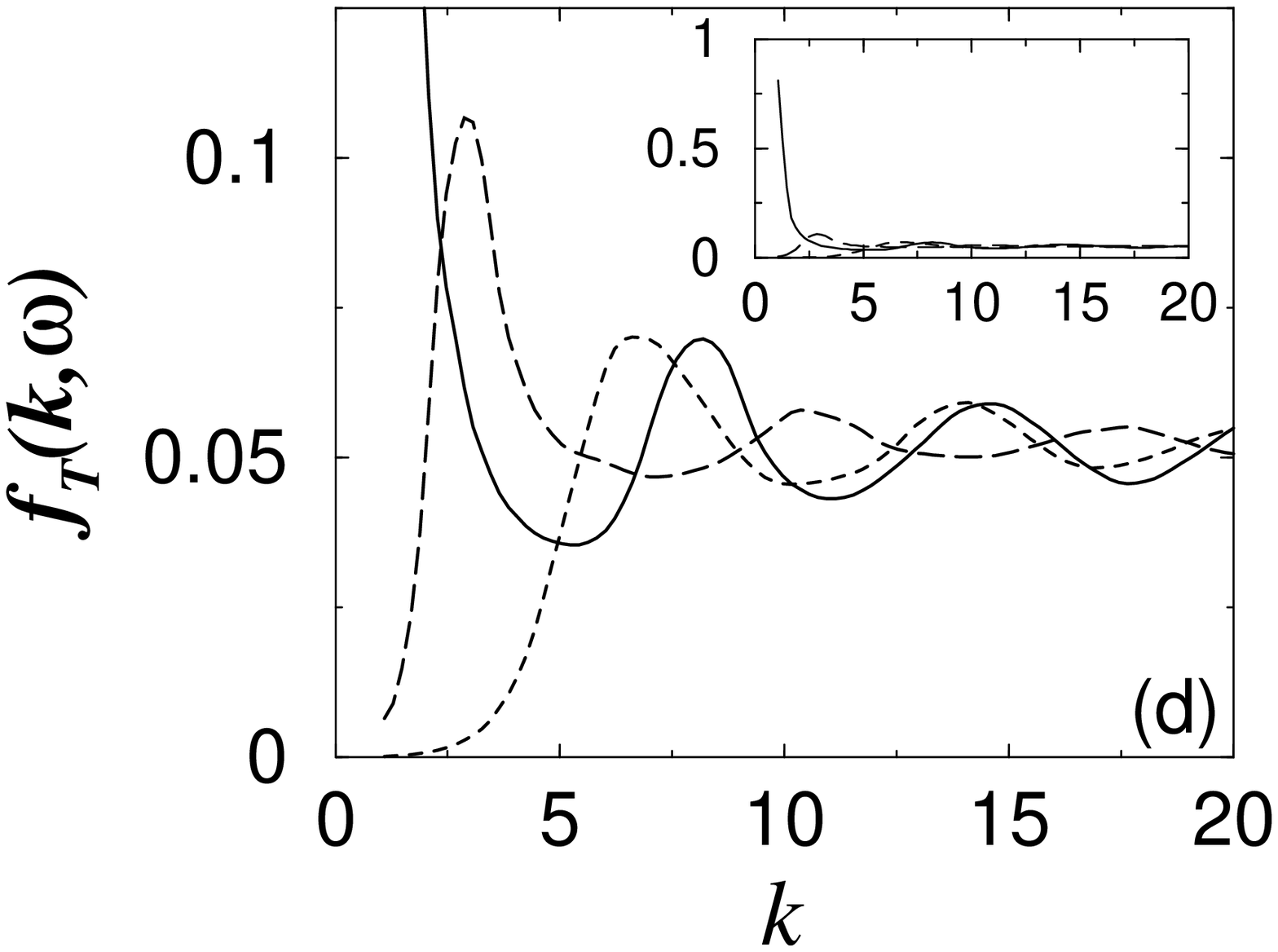}
  \caption{Fourier transforms of the eigenmodes for the low (solid
    line), middle (dashed line), and high frequency (dotted line)
    regions of the vibrational spectrum, at two extreme compressions in 3D.
    Top panels: $\Delta\phi = 1\times 10^{-6}$, for the, (a)
    longitudinal, and (b) the transverse components.  Bottom panels:
    $\Delta\phi = 1\times 10^{-1}$, for the, (c) longitudinal, and (d)
    the transverse components. Insets to (b) and (d) show the
    dominance of the low-$k$ peak in the transverse functions.}
  \label{fig8}
\end{figure}

In a perfect crystal, these functions would be delta-functions at the
wavevectors $k$ in each Brillouin zone characterizing the longitudinal
or transverse vibrational modes at frequency $\omega$. In
Fig.~\ref{fig8}, we show $f_{L}(k,\omega)$ and $f_{T}(k,\omega)$
curves for two disordered configurations in three frequency bands: (i)
at the lowest frequency, (ii) in the middle of the band, and (iii) at
the high-frequency end of the spectrum.  Each curve is averaged over a
narrow bin of frequencies.  For comparison, Figs.~\ref{fig8}(a),(b)
are from a system very close to the jamming threshold, at $\Delta \phi
= 10^{-6}$, and Fig.~\ref{fig8}(c),(d) are for a system that is highly
compressed and far from the jamming threshold, at $\Delta \phi =
10^{-1}$.  At $\Delta \phi=10^{-6}$, the low and mid-frequency curves
correspond to vibrational states in regime B' defined in
Fig.~\ref{fig1}(a), where the density of states is relatively flat,
while the high frequency curve corresponds to regime C. At high
compression, $\Delta \phi = 10^{-1}$, the low frequency curve
corresponds to a state in regime A, the mid-frequency curve
corresponds to regime B and the high frequency curve corresponds to
regime C. The longitudinal functions in general show much more
pronounced structure than do their transverse counterparts. The only
exception to this occurs at very low frequencies and small
wavevectors. In this region the transverse function has a very tall
first peak and the longitudinal function shows only a hint of
structure. The low-wavevector part of the peak in
$f_{T}(k,\omega_{\rm{low}})$ is not resolved because it occurs at $k <
2\pi/L$ where $L$ is the size of the system. That is, the peak is cut
off because of the finite size of the system. The first peak in
$f_{L}(k,\omega_{\rm{low}})$ is also absent for the same reason. In
order to see this structure, one would have to either use a larger box
size at the same value of frequency or else look at $f_{L}(k,\omega)$
at a larger value of $\omega$.

There are multiple oscillations visible in the longitudinal response,
$f_{L}(k,\omega)$. This structure can be thought of as the equivalent
of the repeated structure seen in the higher Brillouin zones of a
crystal \cite{nagel11}. It reflects the large, sharp first peak in the
pair-correlation function, $g(r)$ \cite{leo17}, which leads to strong
oscillations in the structure factor $S(k)=\langle |\sum_{i}
\exp(\imath {\bf k} \cdot {\bf r}) |^{2} \rangle$ \cite{ohern3}.
Similar oscillations show up but with a much smaller amplitude in
$f_{T}(k,\omega)$.

Overall, the results look fairly similar for the two compressions.
For the longitudinal response, the peaks are somewhat smaller and
wider at low compression.  However, the most obvious difference is not
in the peaks but in the minima between them which become more shallow
at small $\Delta \phi$. This is particularly apparent at the lower
frequencies in the longitudinal response. This means that more
wavevectors are making significant contributions to the low-frequency
longitudinal modes at low compression (which are in regime B) than to
low-frequency longitudinal modes at high compression (which are in
regime A), making the mode very different from any plane wave with a
single wavevector. A similar trend is apparent in the transverse
response. Contributions of wavevectors different from the peak value
are relatively larger at low compressions; that is, more wavevectors
contribute to eigenmodes in regime B than in regime A. The
intermediate wavevector oscillations, clearly visible at $\Delta \phi
= 10^{-1}$, are much less pronounced near the onset of jamming. At low
frequency, the transverse response is practically flat at wavevectors
$k > 5$ -- all of these high wavevectors contribute nearly equally in
regime B.

The velocity of longitudinal or transverse sound can be estimated from
the frequency-dependence of the position of the first peak in
$f_{L}(k,\omega)$ or $f_{T}(k,\omega)$. Unlike in crystals, the
dispersion curve is not well-defined because, as we have seen in
Fig.~\ref{fig8}, $f_{L}(k,\omega)$ and $f_{T}(k,\omega)$ have
significant amplitude over a large range of wavevector at all
frequencies. As a result, it is not sufficient to simply plot the
positions of the peaks of the dynamic structure factors as one might
do for a crystal.

\begin{figure}[h!]
  \includegraphics[width=4cm]{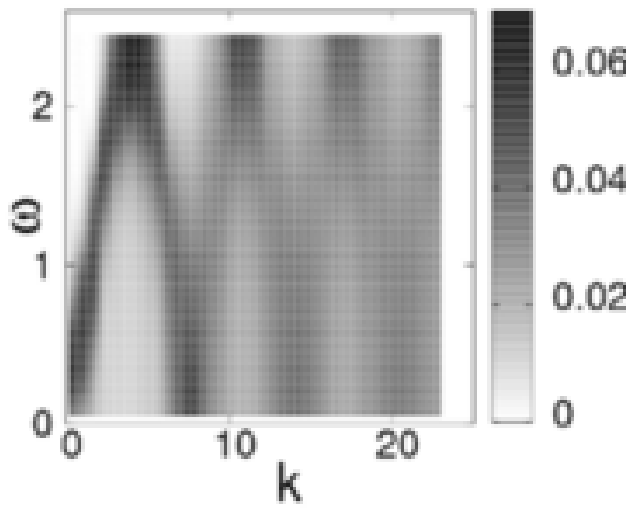}
  \includegraphics[width=4cm]{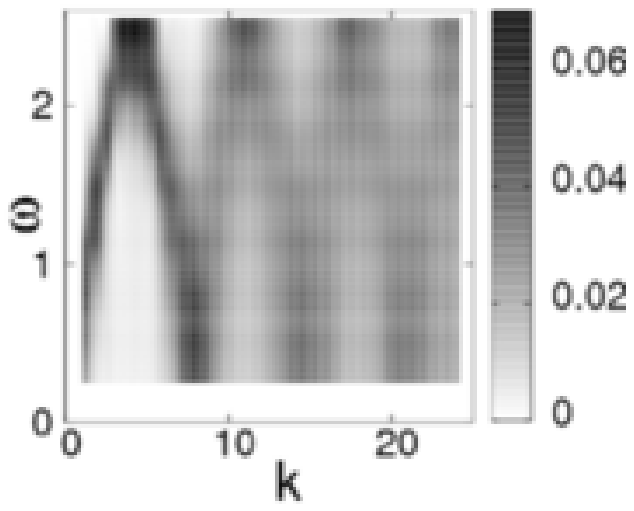}\\
  \includegraphics[width=4cm]{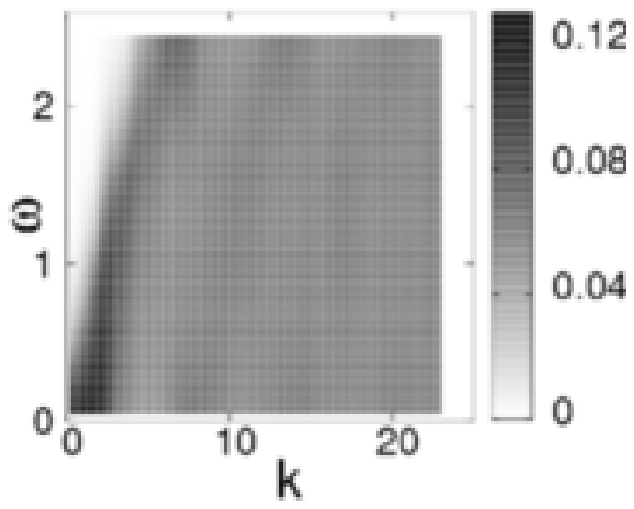}
  \includegraphics[width=4cm]{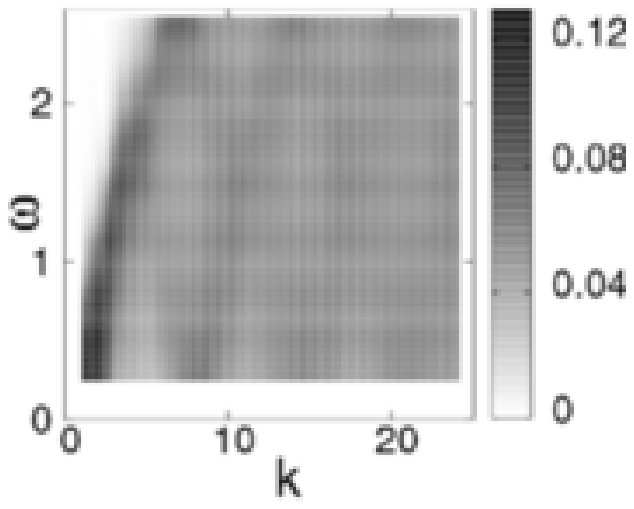}
  \caption{Dispersion relation at two different compressions in 3D. Left
    panels: $\Delta\phi = 1\times 10^{-6}$. Right panels: $\Delta\phi
    = 1\times 10^{-1}$. Top and bottom panels correspond to
    longitudinal and transverse modes respectively. Darker shading
    corresponds to larger amplitude in $f_{L,T}$. The maximum of the
    transverse amplitudes are typically a factor of 5 larger than the
    longitudinal data. The bars to the right indicate the amplitude
    scale.}
  \label{fig9}
\end{figure}

\begin{figure}[h]
  \includegraphics[width=4cm]{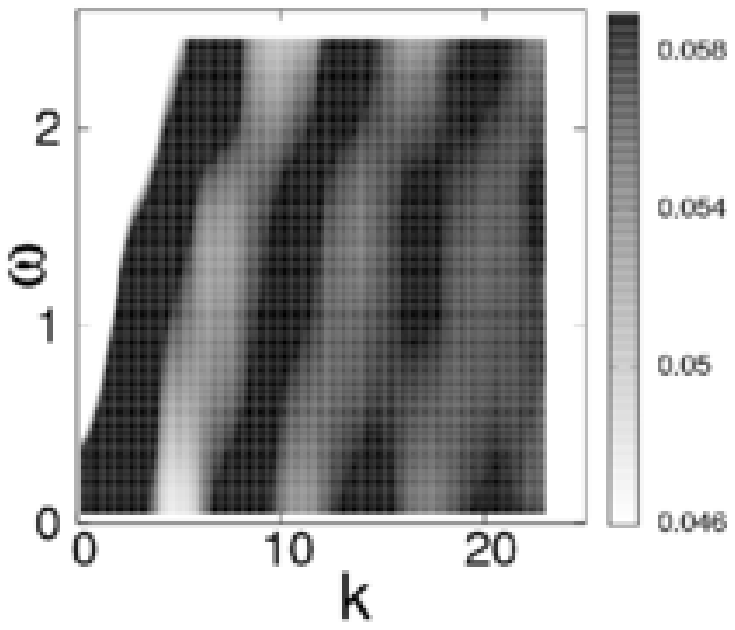}
  \includegraphics[width=4cm]{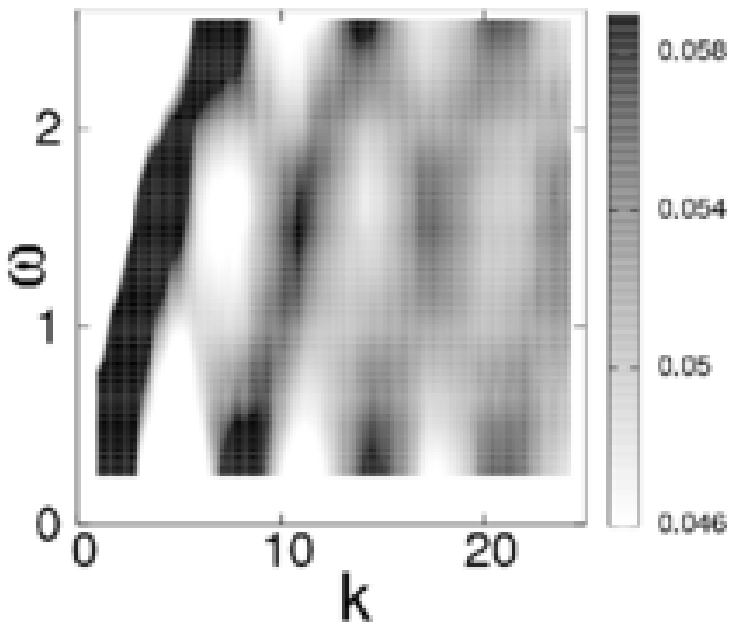}
  \caption{Transverse dispersion data in 3D for $\Delta\phi = 1\times
    10^{-6}$ (left) and $1\times 10^{-1}$ (right), over a limited
    range in amplitude $f_{T}$ (see scale bar).}
  \label{fig10}
\end{figure}

To emphasize the idea of mode mixing, the dispersion data are best
visualized on a gray scale plot as shown in Fig.~\ref{fig9}. (The
features outlined here have been seen in a range of glassy systems,
including soft-sphere \cite{schober2}, covalent \cite{pilla1}, and
metallic \cite{hafner1} glasses.)
Because the amplitude of the first peak in $f_{T}(k,\omega)$ is so
strong, variations at wavevectors greater than the first peak are lost
in Fig.~\ref{fig9}. Therefore, to observe the underlying structure in
the dispersion data at larger wavevectors, in Fig.~\ref{fig10}, we show
the transverse dispersion curves over a limited range in amplitude.
These nicely contrast the unusual underlying structure to the
dispersion relations at the two compressions.

It is clear from looking at these plots that the dispersion relations
are very broad indeed, especially at low compression. Moreover, there
is very little difference between the two compressions, although the
contrast decreases for both transverse and longitudinal modes with
decreasing compression. The limit of what we can achieve is $\Delta
\phi = 10^{-6}$. It is not clear to us if one were able to go even
closer to $\Delta \phi = 0$ whether the variations in
$f_{T}(k,\omega)$ would disappear entirely.

\begin{figure}[t!h]
  \bigskip
  \includegraphics[width=8cm]{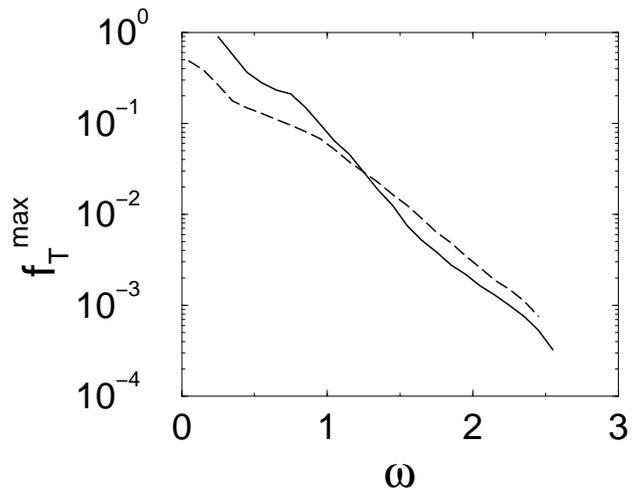}
  \caption{Decay of the peak maximum of the transverse mode transforms
    within the first ``Brillouin zone'', for $\Delta\phi=10^{-1}$
    (solid line) and $10^{-6}$ (dashed line) in 3D.}
  \label{fig11}
\end{figure}

It is also unclear how to define a proper velocity of sound not only
because the peaks in the dispersion relations are so broad, but also
because their amplitudes decay rapidly with increasing $(k,\omega)$.
We illustrate this latter point in Fig.~\ref{fig11} where we show the
approximately exponential decay of the maximum peak height of the
transverse Fourier modes with increasing frequency. We have been
unable to determine precisely whether the decay constant depends on
compression; although there appears to be a small difference between
the two compressions shown in Fig.~\ref{fig11}, that difference is
small.

In summary, we have studied the characteristics of vibrational modes
in different frequency regimes. From the density of states, we see
that there are three regimes: regime A, where the density of states
drops towards zero with vanishing frequency; regime B, where the
density of states is approximately flat; and regime C, where the
density of states is decreasing towards zero with increasing
frequency. As the system is decompressed towards the marginally jammed
state, regime B (or B') increases at the expense of regime A while regime C is
relatively unaffected.  Modes from regime C are localized while modes
from regime A and B are extended.  Those in Regime A are somewhat more
plane-wave-like, in that contributions from wavevectors away from the
peak of the dynamic structure factor are fairly small (the peak is
relatively narrow) while those in Regime B have broader peaks in the
dynamic structure factors, with significant contributions from a wider
range of wavevectors.  Generally, however, we do not observe strong
differences between modes from regime B and regime A.  Thus, the
change in the nature of the modes with decreasing compression is much
less dramatic than the change in the density of vibrational states.

\acknowledgments 

We thank T. Witten for helpful discussions.  We gratefully acknowledge
the support of NSF-DMR-0087349 (AJL), NSF MRSEC DMR-0213745 (SRN),
DE-FG02-03ER46087 AJL,LES), and DE-FG02-03ER46088 (SRN,LES).


\begin{thebibliography}{24}
\bibitem[{wim()}]{wim1}
\bibinfo{note}{We have recently discovered that the distribution of level
  spacings has also been calculated by Z. Zeravcic, W. van Saarloos, and D. R.
  Nelson (unpublished)}.
\end{thebibliography}
\end{document}